# Atomic magnetometry using a metasurface polarizing beamsplitter in silicon on sapphire

Xuting Yang[1], Pritha Mukherjee[1], Minjeong Kim[1], Hongyan Mei[1], Chengyu Fang[1], Soyeon Choi[1], Yuhan Tong[1], Sarah Perlowski[1], David A. Czaplewski[2], Alan M. Dibos[2], Mikhail A. Kats[1], and Jennifer T. Choy[1]

[1]*University of Wisconsin – Madison, Madison, Wisconsin USA*
[2]*Nanoscience and Technology Division, Argonne National Laboratory, Lemont, Illinois USA*

Please send correspondences to jennifer.choy@wisc.edu

**Abstract**

We demonstrate atomic magnetometry using a metasurface polarizing beamsplitter fabricated on a silicon-on-sapphire (SOS) platform. The metasurface splits a beam that is near-resonant with the rubidium atoms (795 nm) into orthogonal linear polarizations, enabling measurement of magnetically sensitive circular birefringence in a rubidium vapor through balanced polarimetry. We incorporated the metasurface into an atomic magnetometer based on nonlinear magneto-optical rotation and measured sub-nanotesla sensitivity, which is limited by low-frequency technical noise and transmission loss through the metasurface. To our knowledge, this work represents the first demonstration of SOS nanophotonics for atom-based sensing and paves the way for highly integrated, miniaturized atomic sensors with enhanced sensitivity and portability.

## 1. Introduction

Atomic magnetometers based on optically pumped atomic vapor are capable of some of the most sensitive measurements of magnetic fields, which are enabled by advances in semiconductor lasers and preparation of atomic spins with long coherence times [1–3]. In combination with advances in miniaturized atomic vapor cell fabrication [2–6], atomic magnetometers are suitable for a host of field applications including biomagnetic imaging in ambient conditions [9–11], navigation and positioning [12–14], and magnetic anomaly detection [15]. In the most general configuration for an optically pumped atomic magnetometer involving a single optical axis, an optical beam nearly resonant with an atomic transition is used to simultaneously polarize atomic spins in an alkali vapor such as rubidium, and probe magnetically sensitive spin rotation within that medium through a polarization change induced on the input beam [8,10,16]. This setup is compatible with miniaturized components for realizing chip-scale atomic magnetometers, but the utilization of bulk birefringent optics for polarization control and measurement in these devices still poses a limit on the ultimate sensor volume and scalability. Therefore, recent research has focused on development of nanophotonic components [17–22] to replace bulk optical elements to achieve a higher degree of miniaturization and integration in atomic magnetometers.

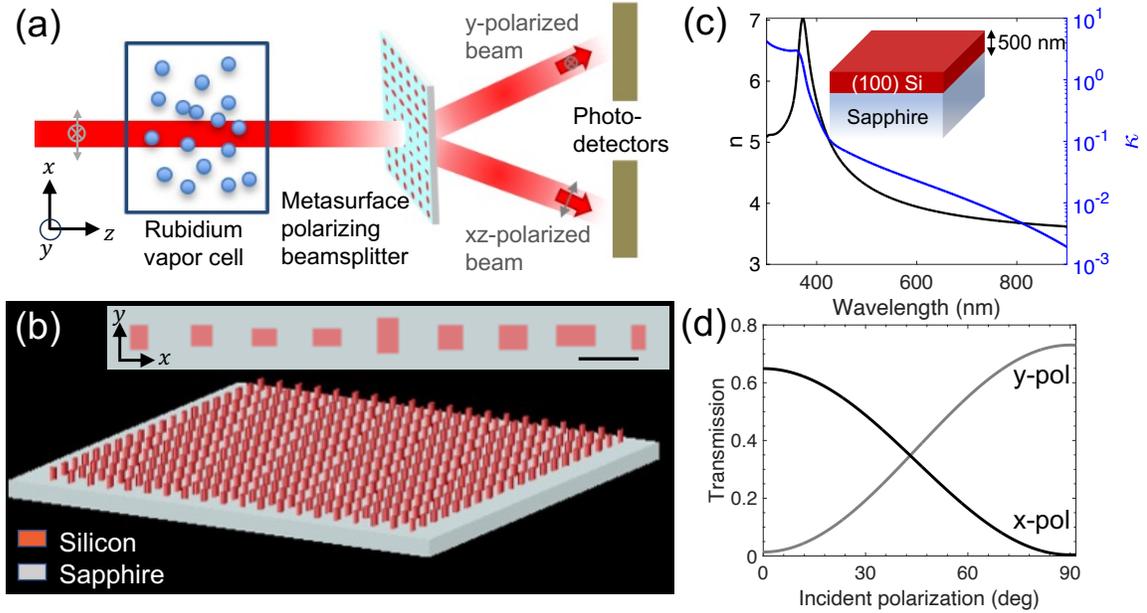

FIG. 1. (a) Optical polarimetry measurement of atomic spins using a transmissive metasurface polarizing beamsplitter (PBS) based on a (b) silicon-on-sapphire (SOS) platform consisting of rectangular meta-atoms in single-crystal silicon on a transparent sapphire substrate. Inset shows the unit cell of the PBS design, which comprises of 9 meta-atoms. The scalebar is 400 nm. (c) Complex refractive index ($n$ and $\kappa$) of silicon extracted using spectroscopic ellipsometry. The thickness of silicon layer is ~500 nm. Details on the ellipsometry are provided in the Supplementary Information SI.I. (d) Simulated transmission into the first-order diffracted modes for the metasurface design.

In this work, we experimentally demonstrate atomic magnetometry with rubidium atoms, using a chip-scale polarizing beamsplitter (PBS) to analyze the atomic spin alignment through balanced polarimetry (Fig. 1a). The PBS is implemented using a dielectric metasurface, which can enable near-complete control of the amplitude, polarization and phase of the light [23–25], fabricated in silicon-on-sapphire (SOS) [26] (Fig. 1b). SOS is a mature material platform widely used in the microelectronics industry [27], due to its excellent thermal stability, electrical insulation, and radiation resistance [28]. The wide optical transparency window of sapphire combined with the high refractive index and moderately low absorption loss of single-crystalline silicon ($n = 3.69$ and $\kappa < 0.01$ at 795 nm, as measured with spectroscopic ellipsometry shown in Fig. 1c) make the SOS platform an attractive candidate for nanophotonic components in the near-infrared (700 nm – 1100 nm), especially devices that do not require propagation over long distances or sharp resonances. The near-infrared wavelength range encompasses many atomic transitions [29,30] relevant for quantum sensing [31,32] and information processing [33,34] applications. In addition, sapphire has a low permeability to gases, is chemically stable, and has been used as a window material in microfabricated silicon alkali metal vapor cells [35–39]. Therefore, SOS nanophotonic components can potentially be pre-patterned on sapphire windows prior to their bonding to vapor cells.

To our knowledge, this work is the first experimental demonstration of the use of SOS nanophotonics in atom-based technologies. We use a metasurface PBS to measure the rotation of rubidium spins and demonstrate its use in a functional atomic magnetometer based on nonlinear magneto-optical rotation (NMOR). The measured sensitivity is below a nanotesla (nT), which is limited by our experimental apparatus and the optical loss through the metasurface. Our metasurface PBS splits the incident light into orthogonal linear polarizations and can be used in combination with a differential photodetector to directly measure the magnetically sensitive circular birefringence in the atomic medium (Fig. 1a). In comparison to recent work demonstrating polarimetry of atomic spins using metasurfaces designed with the Pancharatnam-Berry phase [18,20] (which separate polarization components into circular polarization bases and can therefore directly measure circular dichroism), our technique is more suitable for cases where the circular birefringence is more magnetically sensitive than circular dichroism, including modulation-free NMOR [40] and spin exchange relaxation-free (SERF) magnetometry [16].

## 2. Design, fabrication, and characterization of metasurface PBS

We have previously described our metasurface design approach elsewhere [21]. In the metasurface PBS, the polarization-dependent response is achieved by anisotropic rectangular meta-atoms of the same height ($h = 500$ nm) and periodicity ($a = 400$ nm), whose lateral dimensions vary to locally modify the transmission and phase distribution of each independent polarization component in the incident beam. The transmittance and phase for a square lattice of rectangular posts were obtained through a series of finite-difference time-domain simulations (Ansys Lumerical FDTD) and shown in the Supplementary Information SI.II. To split the polarization of a normally incident beam propagating in $z$, the diffraction angles for the orthogonal polarizations are set to have opposite signs, akin to a blazed grating with oppositely directed blazed angles for the two orthogonal polarizations. We set the phases to be $\varphi_x = -\frac{2\pi}{\lambda} x \sin\theta$ for the $x$-polarized component and $\varphi_y = +\frac{2\pi}{\lambda} x \sin\theta$ for the $y$-polarized component. We select $2\theta = 25.5°$ to be the split angle and $\lambda = 795$ nm for operation near the rubidium (Rb) D1 transition. The metasurface is then constructed by selecting geometric parameters of the meta-atoms that match this profile while minimizing the phase matching error $\epsilon = |t_x - e^{i\varphi_x}|^2 + |t_y - e^{i\varphi_y}|^2$ where $t_{x,y}$ is the polarization-dependent complex transmission amplitude of the meta-atom [24]. Our design is translationally invariant along $y$ and consists of a unit cell comprising 9 meta-atoms that is repeated along $x$ (Fig. 1b). As shown in Fig. 1d, our metasurface PBS design exhibits polarization-dependent beam-splitting, with the polarization extinction ratio (PER, defined as the ratio of the transmitted powers into the split beams for each incident $x$ or $y$ polarization component) calculated to be 16.7 dB (for $x$-polarized input) and 22.6 dB (for $y$-polarized input). The total transmission efficiency of the simulated structure is over 70%, with roughly 20% of the incident light reflected and 6-7% absorbed by the structure (see Supplementary Information SI.III).

Our fabrication procedure is illustrated in Fig. 2a. The metasurface PBS was fabricated on an SOS wafer with the silicon thickness $h = 500$ nm and resistivity > 100 Ω·cm. A 50-nm silicon dioxide ($SiO_2$) hard mask was first deposited on the silicon using plasma-enhanced chemical vapor deposition (PECVD), followed by spin-coating of a positive electron-beam resist (ZEP 520A). The metasurface pattern was written using electron-beam lithography at a beam current of 4 nA and patterned hard mask was subsequently formed by etching of the $SiO_2$ layer in a mixture of $CHF_3$ and $O_2$ gases. Finally, a mixture of HBr and $O_2$ gases was used to etch the silicon metasurface structure using inductively coupled plasma reactive ion etching (ICP-RIE). While the hard etch mask can be removed with a buffered oxide etch, based on simulations the remaining $SiO_2$ mask does not meaningfully affect the metasurface performance (Supplementary Information SI.V).

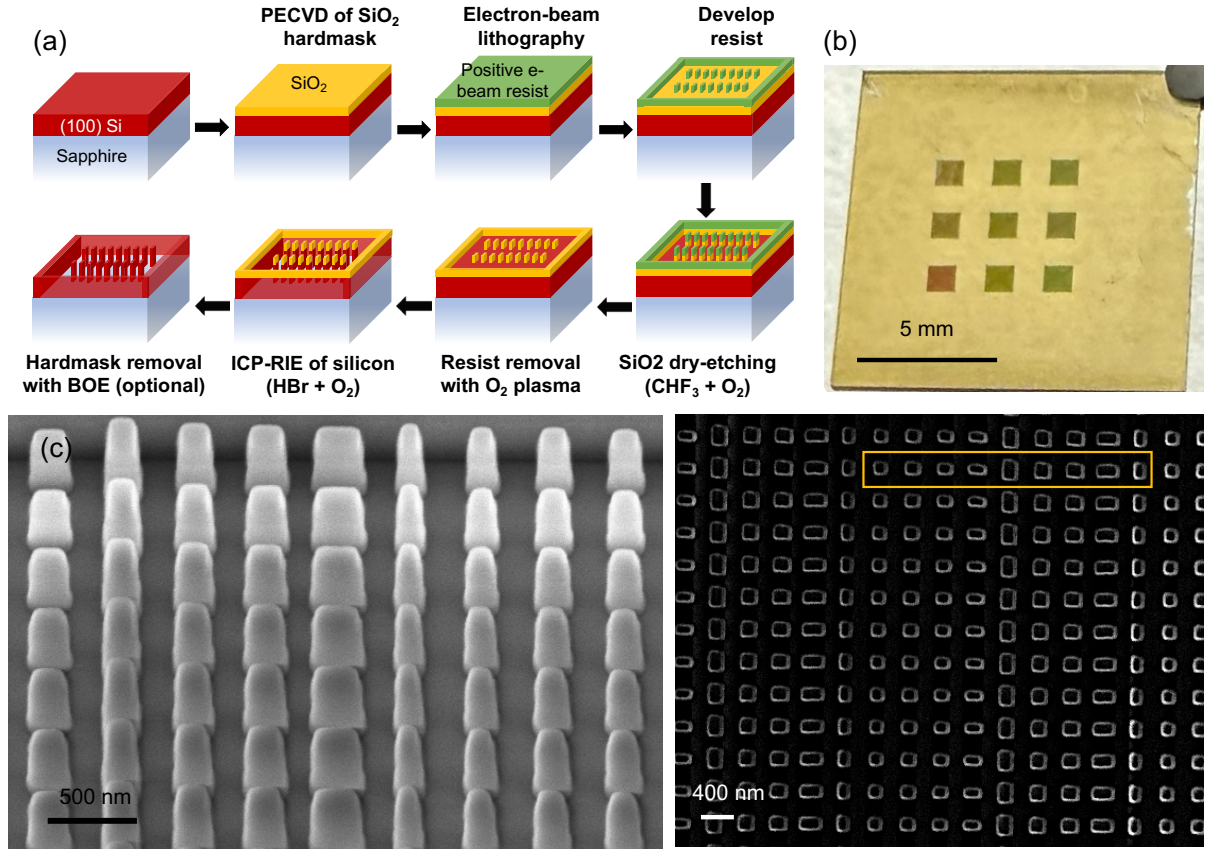

FIG. 2. (a) Schematic of the fabrication process. BOE: buffered oxide etch. (b) Photo of arrays of metasurface PBS (each 1 mm by 1 mm) fabricated on an SOS sample. (c) Side profile (taken at 30 degrees) and top-down SEM images of the etched silicon posts. For the metasurface array shown, a unit cell is highlighted.

The SOS chip after fabrication can be seen in Fig. 2b, with each 1 mm² array of etched silicon metasurfaces consisting of a different dose within the 3 × 3 array. Figs. 2c show the scanning electron microscopy (SEM) images of the metasurface PBS. For the fabricated samples in this work, the silicon layer is estimated to be under-etched by ~10 nm and the sidewall profile for each meta-atom is slightly tapered with the angle estimated to be 7° ± 2°. Both these effects can be attributed to drifts in the silicon etch anisotropy and etch rate due to charging in the silicon and sapphire, which might be mitigated with the use of doped silicon in the future.

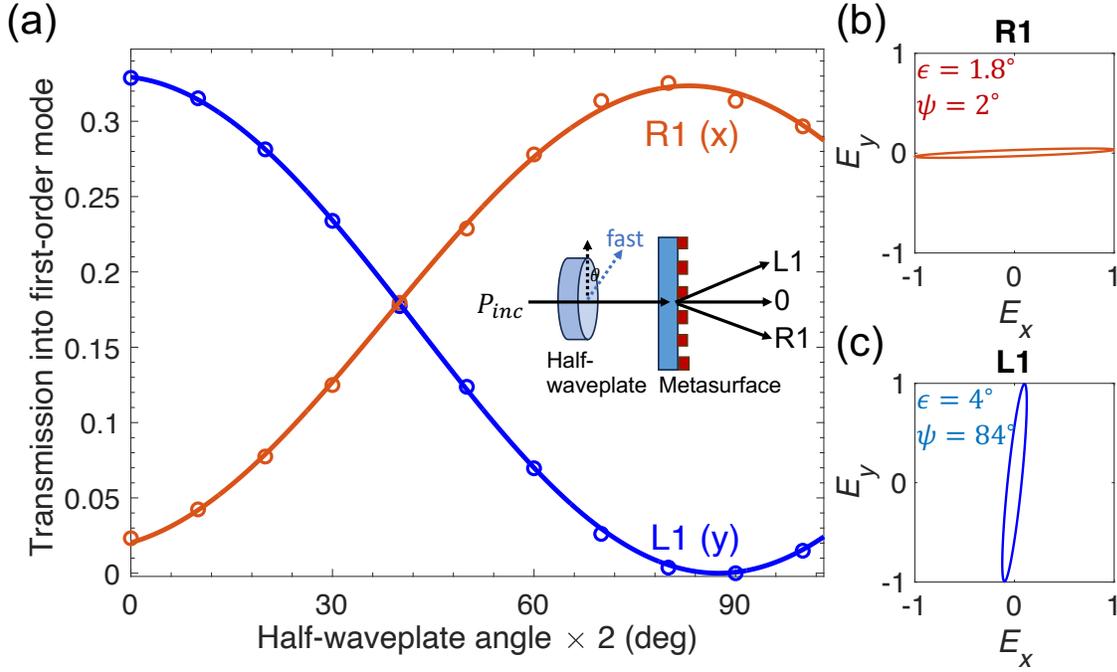

FIG. 3 (a) Measured transmission (transmitted power divided by the incident power $P_{inc}$) into each of the polarization-dependent modes as a function of incident polarization. The inset shows the labeling of the zeroth order ("0") and diffracted modes in the transmitted beam. (b)-(c) Polarization ellipses for the first-order diffracted modes, along with the ellipticity ($\epsilon$) and azimuth ($\psi$) values, taken at an incident polarization that yielded balanced outputs at R1 and R1.

The optical performance of the metasurface was measured using a linearly polarized 795-nm distributed Bragg reflector (DBR) laser close to normal incidence on the device, with the incident polarization angle varied using a half waveplate. The sample was mounted on a tip-tilt stage which allows for the alignment to be adjusted to optimize the balancing of transmission into the split beams. We observed polarization-dependent splitting, shown in Fig. 3a, of the transmitted beam into the first-order diffracted beams, which are labeled R1 (for the *x*-polarized incident beam) and L1 (for the *y*-polarized incident beam). The maximum transmission into each port (normalized to the incident power) is 33%. The PER is determined to be $PER_x = 460$ (26.6 dB) and $PER_y = 16$ (12 dB). The output polarization into each first-order mode was measured on a polarimeter observed to largely correspond to the respective incident $E_x$ and $E_y$ electric field components (Fig. 3b and c). The slight ellipticity introduced in the transmitted beams and difference in PER between the two ports can be attributed to linear dichroism in the structure.

The overall transmission through the structure is 59%, with significant reflection (not quantified) observed in agreement with simulations (Supplementary Information SI.III). This reflection may be reduced by incorporating anti-reflection coating in the future. In addition to the first-order modes, the transmitted light was distributed into zeroth-order mode (accounting for 11.8% of the signal) and additional higher-order modes amounting to few percent of the total incident power (see Supplementary Information SI.IV for the normalized transmitted powers at the balanced point between the two first-order modes). The occurrence of the zeroth-order and higher-order modes can be attributed to the tapered profile of the silicon meta-atoms, based on simulations (Supplementary Information SI.V). Whereas the designed geometry suppresses transmission into the zeroth- and higher-order modes so that over 95% of the transmitted light goes into the first-order modes, any deviation from a straight sidewall causes the incident power to be distributed into these undesired channels.

## 3. Nonlinear magneto-optic resonance measurements using the metasurface PBS

When near-resonant light interacts with an atomic medium in a magnetic field applied along the light-propagation direction, an optical-power-dependent polarization rotation is observed [41]. This phenomenon is called non-linear magneto-optical rotation (NMOR). Compared to its linear counterpart, NMOR can exhibit extremely narrow resonance linewidth, making it a promising candidate for precision magnetometry [42].

We incorporated our metasurface PBS into atomic magnetometry based on NMOR, since the simple measurement configuration (requiring a linearly polarized input beam and vapor cell operating at near room temperature) facilitates proof-of-principle demonstration of atomic spin detection using a chip-scale element. The experimental setup for the NMOR atomic magnetometry is shown in Fig. 4a. A DBR laser was frequency locked to the F = 2→F' = 1 transition of the D1 line of $^{87}$Rb (Fig 4b) via saturated-absorption spectroscopy. The output from the polarization-maintaining fiber was passed through a linear polarizer (PER > $10^5$) with its intensity actively stabilized through a noise eater (Thorlabs NEL03A). A half waveplate set the input polarization of the laser before going through the isotopically pure $^{87}$Rb vapor cell housed inside a four-layer magnetic shield (Twinleaf MS-1L). A low-noise bipolar current supply (Twinleaf CSBA) drove the magnetic coil inside the shield for precise control of the magnetic field in the center of the shielded volume. The current was adjusted either manually to produce static magnetic fields between −4 µT and +4 µT (Fig. 4c) or by a slow sweep around the zero-field point (Fig. 4d). For field calibration during the measurements with manual tuning, a 3-axis magnetoresistive probe (AlphaLab MR3) was co-located with the vapor cell and gave real-time reading of the magnetic field. The cell (25 mm in diameter and 75 mm in length) was coated with paraffin to extend the ground state polarization lifetime and is maintained at a temperature of 50 °C by an alternating current heat controller (Twinleaf TCHF) to minimize the stray DC field from the heating tape attached to the cell. The magnetometry signal was detected by a balanced polarimeter incorporating a half waveplate, a metasurface PBS and a balanced photodetector (Thorlabs PDB210A). The differential output from the balanced detector was collected and analyzed by a National Instrument multifunction I/O device (PCIe 6363). Each data point required an integration time of 10s at a sample rate of $10^6$ samples/s.

Figure 4c shows the differential voltage output from the balanced detector under different optical powers as the longitudinal field is manually swept across zero. The curves all show magnetic sensitivity in the differential voltage near zero field ($|B_z| < 100$ nT). At each input power, the sharp slope observed near the zero-field point is attributed to the paraffin coating preserving the atomic alignment over thousands of atom-wall collisions [43,41]. As predicted by the calculation in [41], a lower light intensity gives rise to a narrower NMOR feature. However, operating under 20 μW was not ideal in our setup due to the limited signal-to-noise ratio (SNR) in the photo-detection. In our setup, we found that $P = 50$ μW (corresponding to $I = 400$ μW/cm$^2$) yielded the best magnetic sensitivity as a compromise between SNR and magnetic linewidth. As the reference magneto-resistive probe measurements became unreliable with longitudinal fields below 5 nT, we performed a finer characterization of the differential signal with respect to small longitudinal field at this light intensity by a slow sweep of the current around zero field. Fig. 4d shows the differential voltage signal swept across a few nT's (with the values of the applied magnetic field calculated using specifications by the coil manufacturer). The linear fit of the differential signal gives a slope of $-1.373$ mV/nT, which translates to a noise-equivalent magnetic field of 0.8660 nT at $B_z = 0$ based on a voltage noise of $\sigma(V_{diff}) = 1.189$ mV (measured by the standard deviation of the differential voltage near zero field).

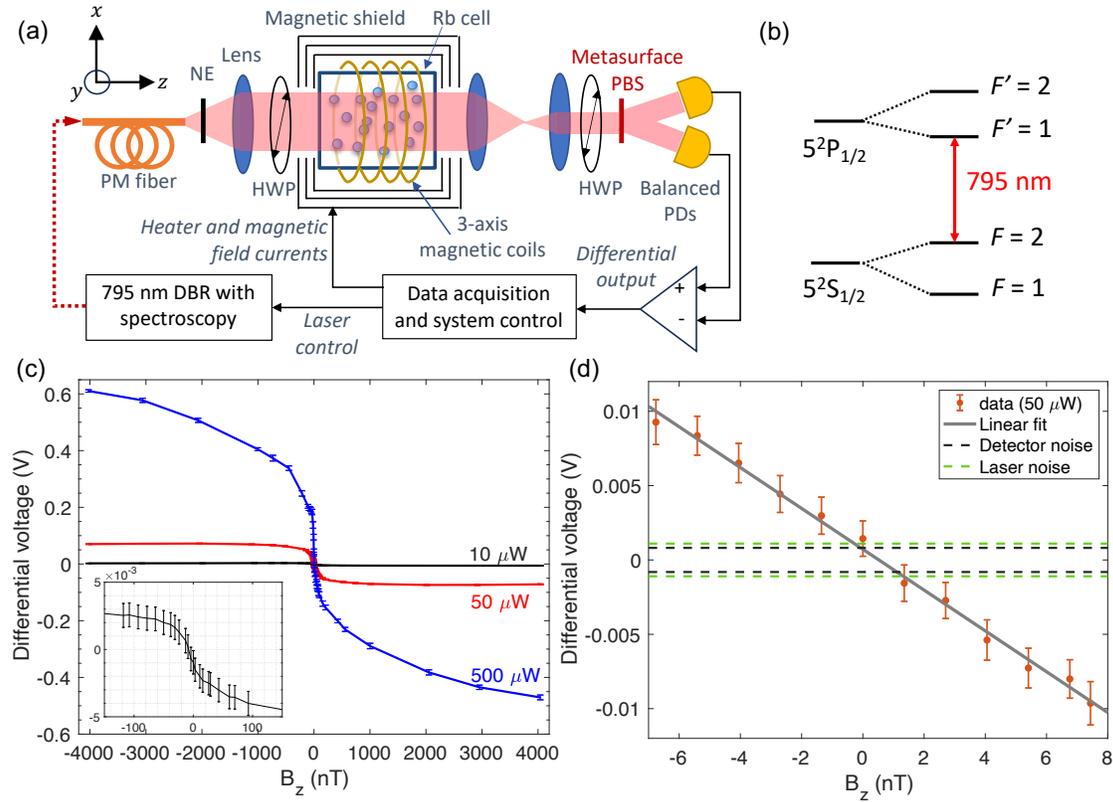

FIG. 4. (a) Schematic of the atomic magnetometry setup in which polarimetry is performed with a metasurface PBS. PM fiber: polarization-maintaining fiber; NE: noise-eater; HWP: half-wave plate; PD: photodiode. (b) Energy level diagram for the NMOR measurement, in which the incident beam is tuned to the $F = 2 \to F' = 1$ transition in $^{87}$Rb. (c)-(d) Differential voltage signal as a function of applied magnetic field, for different optical powers. The inset in (c) shows the signal around zero field for a pump power of 10 μW, while (d) shows the signal around zero field for a pump power of 50 μW. The error bars at each data point indicate the standard deviations in the recorded time trace, while the dotted lines represent the standard deviations of the non-magnetic noise contributions from the detector and laser.

We further characterized the technical noise contributions to the magnetometer system by taking measurements near $B_z = 0$. To decouple the laser noise from the magnetic noise, we detuned the laser far from any atomic transitions of rubidium by $> 10$ GHz, where both the linear and non-linear magneto-optical effects are small. With laser completely blocked, we measured a standard deviation of 0.8449 mV for output signals, a noise in agreement with the technical specification for the photodetector. The voltage noise slightly increased to 1.111 mV with an incident laser power of 50 µW. The results indicate that the detection is limited mainly by photodetector noise and laser noise. Therefore, increasing the polarization splitting efficiency of the metasurface will directly improve the SNR and boost the magnetic sensitivity. Other possible improvements include using a higher-quality coating [44] for increased ground state alignment lifetime, performing detection at frequencies where technical noises are suppressed through modulation techniques [39], and varying the transverse field to obtain a sharper magnetic resonance feature [45].

## 4. Conclusion

We have successfully demonstrated rubidium atomic magnetometry using an SOS metasurface PBS operating in the linear polarization basis. By measuring the differential voltage signal proportional to the optical rotation without any phase sensitive (lock-in) detection, we directly measured the magnetometer sensitivity using the outputs of the metasurface PBS. Our experiment shows the effectiveness of the metasurface PBS in magnetometry signal detection and paves the way for nanophotonic integration of atomic magnetometers on an SOS platform.

The magnetometry scheme used for this demonstration is based on DC NMOR near zero field and thus prone to low-frequency noise especially without lock-in detection. Therefore, our sensitivity is limited by low-frequency photodetector and laser noise. In prior work, we showed that the detected differential signal under an imperfect PBS (for example, one that demonstrates linear dichroism and transmission loss) is attenuated by a coefficient $\xi < 1$ that depends on its transmission and PER [21]. Therefore, a magnetometer system dominated by the photodetector noise will benefit from an improvement in the metasurface PBS performance by increasing $\xi$ and thus gaining higher SNR. Furthermore, an increase in the optical rotation signal allows one to operate in a lower optical power region, which can lead to a narrower magnetic resonance linewidth. In the near-term, the transmission efficiency of the metasurface PBS may be improved by incorporating antireflection coatings and the use of doped silicon to mitigate fabrication imperfections caused by charging. Additionally, the device area, limited to the order of mm$^2$ in this work, can be significantly increased by using 193 nm immersion lithography [46,47] whose resolutions are compatible with our metasurface design.

Our work shows that SOS is a powerful platform for integrating nanophotonics with atomic sensing. The high transparency in the near-infrared and ultra-low gas permeability in the sapphire substrate are well suited for integration with miniaturized rubidium or cesium vapor cells, while the high refractive index silicon enables efficient light manipulation, resulting in highly integrated atomic sensing platforms. Together with recent progress in photonic integrated circuits and laser technology, chip-scale atomic sensors with exceptional sensitivity, low footprint and high portability may be feasible in the future.

**Acknowledgments**

This material is based upon work supported by the Office of Naval Research under Grant No. N00014-20-1-2598 and the U.S. Department of Energy Office of Science National Quantum Information Science Research Centers. PBS nanofabrication was performed at the Center for Nanoscale Materials, a U.S. Department of Energy Office of Science User Facility, was supported by the U.S. DOE, Office of Basic Energy Sciences, under Contract No. DE-AC02-06CH11357.

The authors are grateful to helpful discussions with Arka Majumdar, who suggested the use of a rectangular geometry (vs. using rounded edges) to reduce lithography time.

**Disclosures** The authors declare no conflicts of interest.

**Data availability** Data underlying the results presented in this paper may be obtained from the authors upon request.

**Supplementary Information**

**SI.I: Extraction of the complex refractive index using spectroscopic ellipsometry**

In order to extract the optical properties of silicon for the SOS wafer from Roditi, we employed generalized spectroscopic ellipsometry (VASE, J.A. Woollam Co.) [48] as well as the transmission/reflection spectroscopy. Figure S1(a, b) shows the raw measured ellipsometry data $\Psi$ and $\Delta$ of $A_{ne}$ (in green) and the corresponding model fitted data (in red) for the center region of the SOS wafer with the angle of incidence $\theta_i = 60°$. $A_{ne}$ is the ratio between the two Fresnel coefficients $r_{pp}$ and $r_{ss}$. R-plane cut sapphire is slightly anisotropic, but from the measurement we found out the cross-polarization terms $A_{ps}$ and $A_{sp}$ is close to zero. We also observe no obvious difference of the fitting results when considering anisotropy of sapphire or assuming sapphire as an isotropic substrate. Additionally, as shown in Figure S2, we measured reflection of the SOS wafer at normal incidence under different polarization of light and observe no shift of the Fabry–Pérot fringes.

$$A_{ne} = \rho = \frac{r_{pp}}{r_{ss}} = \tan(\Psi_{A_{ne}}) e^{i\Delta_{A_{ne}}}$$

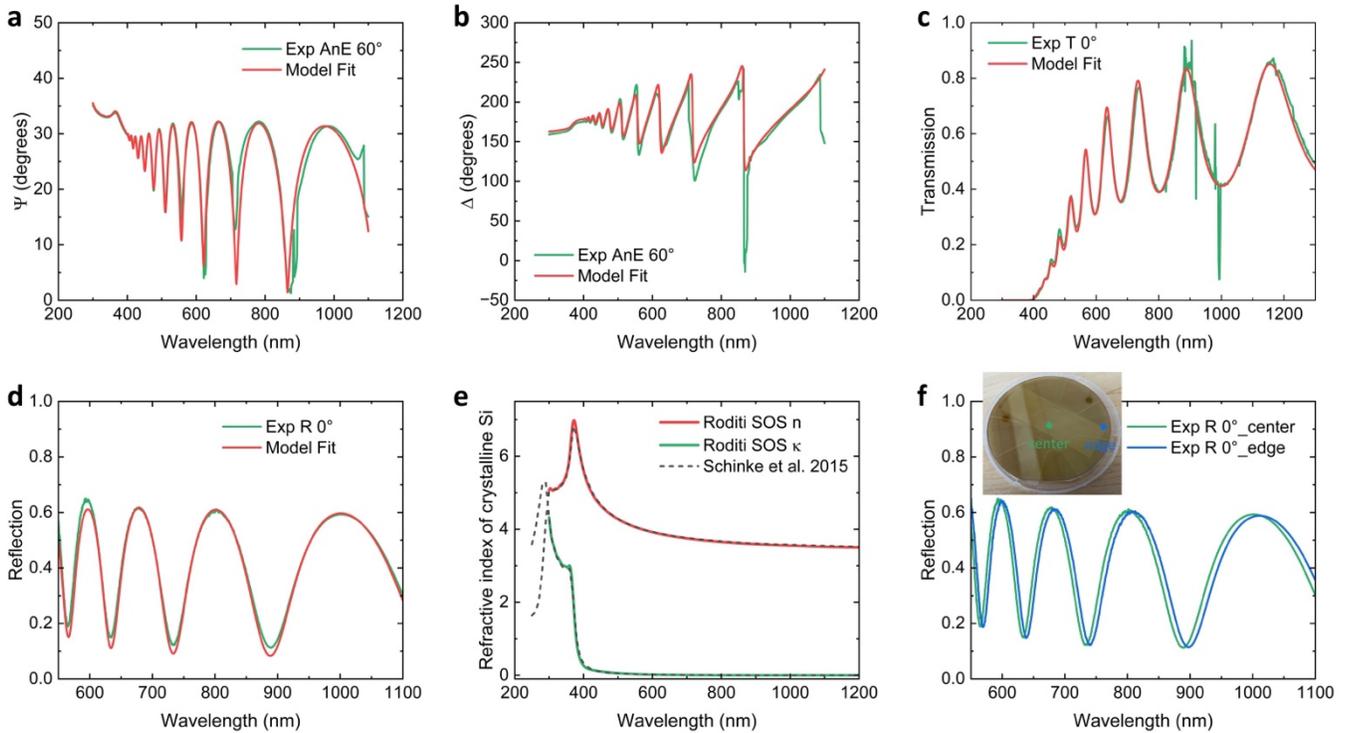

**Figure S1. (a-d)** Raw ellipsometry data $\Psi$ **(a)** and $\Delta$ **(b)** of $A_{ne}$ for silicon on sapphire wafer at $\theta_i = 60°$, transmission **(c)** and reflection **(d)** data at normal incidence, showing good consistence between the experimental data (green lines) and model fit data (red lines). **(e)** Extracted complex refractive index of single crystalline silicon of Roditi SOS wafer (solid lines), compared with single crystalline silicon data from literature[2] (dashed lines). **(f)** Measured reflection spectra of the SOS wafer at the center and edge, the shift of the fringes showing the variation of thickness.

$$A_{ps} = \frac{r_{ps}}{r_{pp}} = \tan\left(\Psi_{A_{ps}}\right) e^{i\Delta_{A_{ps}}}$$

$$A_{sp} = \frac{r_{sp}}{r_{ss}} = \tan\left(\Psi_{A_{sp}}\right) e^{i\Delta_{A_{sp}}}$$

We also included the transmission and reflection spectra into the model for fitting, and the extracted complex refractive index of silicon is shown in Figure S1(e), showing nice consistency with the data from recent literature [49]. In addition, we examined different regions across the 4-inch wafer and found that there are small variations of the thickness of the silicon layer. The shift of the Fabry–Pérot fringes in Figure S1(f) clearly reveal that. According the fitting, we observe the thickness of the silicon layer is $494.0 \pm 0.232$ nm at the center, and $499.2 \pm 0.076$ nm at the edge of the wafer.

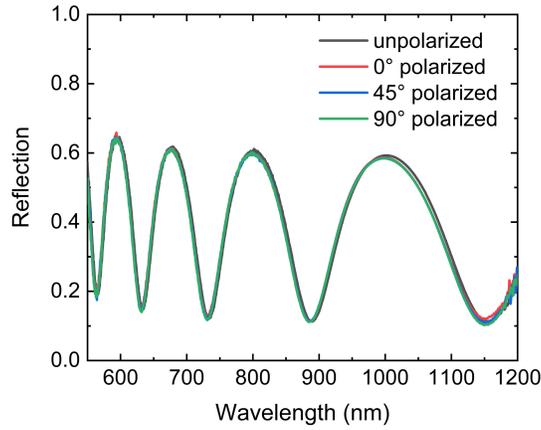

**Figure S2.** Measured reflection spectra of the SOS wafer with normal incident light unpolarized, and polarized at different angles, showing no obvious difference.

**SI.II: Phase shift and transmission amplitude of the rectangular meta-atoms as a function of the length $l_x$ and width $l_y$**

We constructed the metasurface by matching the designed birefringent phase profile with the phase and transmission response of the rectangular meta-atoms, which were calculated with FDTD for a square lattice of silicon rectangular rods in air with a fixed height of 500 nm and periodicity 400 nm. The measured $(n, k)$ data described in SI I was used for the simulation. Figure S3 shows the phase and transmission response of the silicon square lattice for different transverse dimensions ($L_x$ and $L_y$). By varying the geometries, the meta-atoms provide a full $2\pi$ phase coverage while maintaining relatively high transmission.

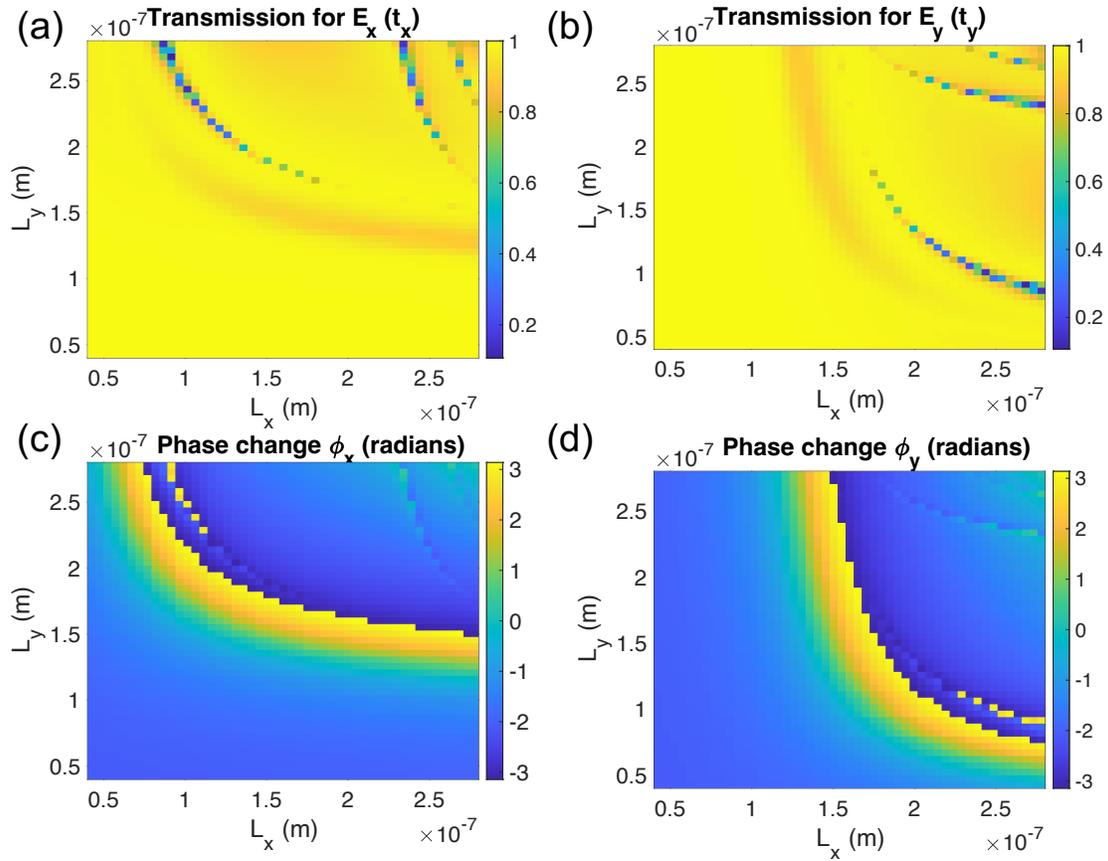

**Figure S3.** Calculated normalized transmission and phase shift of an incident x- or y- polarized plane wave, based on full-wave simulation of a square silicon array of rectangular rods with fixed lattice constant (a = 400 nm), post height (h = 500 nm), and swept transverse dimensions $L_x$ and $L_y$.

**SI.III: Calculated transmission and reflection of the metasurface PBS design**

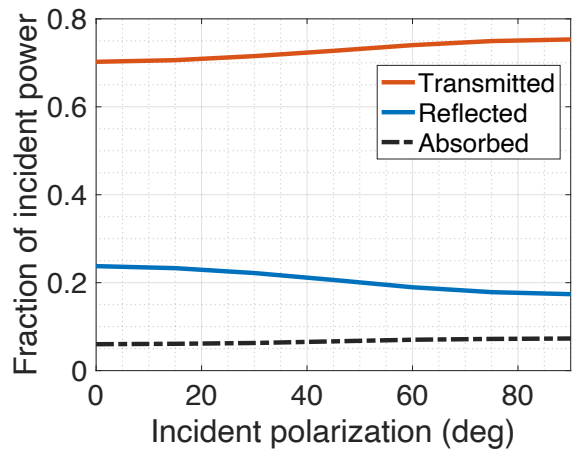

**Figure S4.** Calculated total transmission, reflection, and absorption (in the device layer) for the metasurface PBS.

**SI.IV: Measurement of the power distribution of the transmitted beam through the metasurface**

Table S1 Transmitted power into the zeroth, first, and second order modes, for an incident beam polarization that best balanced the outputs into R1 and L1.

| Transmitted beam | Fraction transmitted (relative to $P_{inc}$) |
|---|---|
| R1 (first-order) | 13.66% |
| L1 (first-order) | 13.50% |
| 0 (zeroth-order) | 11.83% |
| R2 (second-order) | 1.92% |
| L2 (second-order) | 3.17% |

**SI.V: Simulation of non-ideal geometries**

We modeled the effects of various deviations from metasurface design due to fabrication, such as the inclusion of a 50-nm-thick $SiO_2$ etch mask, a remaining silicon pedestal layer from under-etching, and tapering in the sidewall profile of the meta-atoms. The calculated transmission and reflection values at an incident polarization of 45° are shown in Table S2. We note that this set of simulations does not account for the reflection of an incident beam at the first air-sapphire interface.

Table S2 Calculated transmission and reflection through the metasurface PBS.

|  | Fraction of incident power | | | | |
|---|---|---|---|---|---|
|  | L1 (y) | R1 (x) | $0^{th}$ order | Total transmitted | Total reflected |
| As designed | 0.40 | 0.35 | < 0.02 | 0.79 | 0.14 |
| With $SiO_2$ etch mask | 0.42 | 0.36 | < 0.02 | 0.81 | 0.13 |
| With 10 nm pedestal | 0.36 | 0.30 | < 0.02 | 0.70 | 0.23 |
| With 15 nm pedestal | 0.32 | 0.29 | < 0.02 | 0.66 | 0.27 |
| With 5° taper | 0.16 | 0.17 | 0.37 | 0.76 | 0.17 |

Simulation results suggest that the $SiO_2$ etch mask does not alter the transmission or beam-splitting performance, while a silicon pedestal layer as thin as 10-15 nm can significantly increase reflection, thereby reducing transmission.

We modeled a tapered sidewall structure based on SEM images of the etched silicon. The simulated optical response of the metasurface PBS with a 5 degree sidewall tapering is plotted in Figure S5 which shows a prominent zeroth-order mode and higher order scattering modes appearing in the far field.

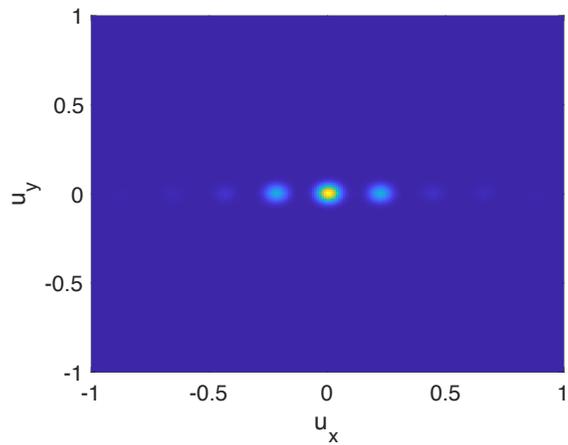

**Figure S5.** Simulated far-field profile of the metasurface design with a 5° taper in the sidewall angle.